\documentclass[%
reprint, hidelinks,
superscriptaddress,
hypertext,
showkeys,
showpacs,
 nofootinbib,
 nobibnotes,
 amsmath,amssymb,
 aps,
prc,
floatfix,
]{revtex4-2}

\usepackage{graphicx}
\usepackage{dcolumn}
\usepackage{bm}
\usepackage{isotope}
\usepackage{upgreek}
\usepackage{longtable}
\usepackage[caption=false]{subfig}
\usepackage{booktabs}
\usepackage{threeparttable}
\usepackage{epigraph}
\usepackage{verbatim}
\usepackage{comment}
\usepackage[]{hyperref}
\hypersetup{
  colorlinks   = true, 
  urlcolor     = magenta, 
  linkcolor    = magenta, 
  citecolor   =  cyan 
}
\usepackage[mathlines]{lineno}

\newcommand{\beag}{$^{7}$Be$(\alpha,\gamma)^{11}$C}

\begin{document}

\preprint{APS/123-QED}

\title{Direct measurement of resonances in $^7$Be($\alpha,\gamma$)$^{11}$C relevant \\ to $\nu p$--process nucleosynthesis}

\author{A. Psaltis}
\email{psaltisa@mcmaster.ca}
\altaffiliation[Present address: ]{Institut f\"ur Kernphysik,
Technische Universit\"at Darmstadt, 64289 Darmstadt, Germany
}
\affiliation{
 Department of Physics \& Astronomy, McMaster University,
 Hamilton, Ontario L8S 4M1, Canada
}
\affiliation{
  The NuGrid collaboration \href{https://nugrid.github.io/}{https://nugrid.github.io/}
}
\author{A.A. Chen}
\affiliation{
 Department of Physics \& Astronomy, McMaster University,
 Hamilton, Ontario L8S 4M1, Canada
}
\affiliation{
 The NuGrid collaboration \href{https://nugrid.github.io/}{https://nugrid.github.io/}
}
\author{R. Longland}
\affiliation{Department of Physics, North Carolina State University,
Raleigh, NC, 27695, USA}
\affiliation{Triangle Universities Nuclear Laboratory, Duke University, Durham, NC, 27710, USA}
\author{D.S. Connolly}
\altaffiliation[Present address: ]{Los Alamos National Laboratory, \\
Los Alamos, NM 87545, USA}
\affiliation{TRIUMF, 4004 Wesbrook Mall, Vancouver,
British Columbia V6T 2A3, Canada}
\author{C.R. Brune}
\affiliation{Department of Physics \& Astronomy,
Ohio University, Athens, Ohio 45701, USA}
\author{B. Davids}
\affiliation{TRIUMF, 4004 Wesbrook Mall, Vancouver,
British Columbia V6T 2A3, Canada}
\affiliation{Department of Physics, Simon Fraser University, Burnaby, British Columbia V5A 1S6, Canada}
\author{J. Fallis}
\affiliation{North Island College, 2300 Ryan Rd, Courtenay,
BC V9N 8N6, Canada}
\author{R. Giri}
\affiliation{Department of Physics \& Astronomy,
Ohio University, Athens, Ohio 45701, USA}
\author{U. Greife}
\affiliation{Department of Physics, Colorado School of Mines,
Golden, Colorado 80401, USA}
\author{D.A. Hutcheon}
\affiliation{TRIUMF, 4004 Wesbrook Mall, Vancouver,
British Columbia V6T 2A3, Canada}
\author{L. Kroll}
\affiliation{
 Department of Physics \& Astronomy, McMaster University,
 Hamilton, Ontario L8S 4M1, Canada
}
\affiliation{
   The NuGrid collaboration \href{https://nugrid.github.io/}{https://nugrid.github.io/}
}
\author{A. Lennarz}
\affiliation{TRIUMF, 4004 Wesbrook Mall, Vancouver,
British Columbia V6T 2A3, Canada}
\author{J. Liang}
\altaffiliation[Present address: ]{TRIUMF, 4004 Wesbrook Mall, Vancouver,
British Columbia V6T 2A3, Canada
}
\affiliation{
 Department of Physics \& Astronomy, McMaster University,
 Hamilton, Ontario L8S 4M1, Canada
}
\author{M. Lovely}
\affiliation{Department of Physics, Colorado School of Mines,
Golden, Colorado 80401, USA}
\author{M. Luo}
\affiliation{Department of Physics \& Astronomy, University of British
Columbia, Vancouver, British Columbia V6T 1Z4, Canada}
\author{C. Marshall}
\altaffiliation[Present address: ]{Department of Physics \& Astronomy,
Ohio University, Athens, Ohio 45701, USA}
\affiliation{Department of Physics, North Carolina State University,
Raleigh, NC, 27695, USA}
\affiliation{Triangle Universities Nuclear Laboratory, Duke University, Durham, NC, 27710, USA}

\author{S.N. Paneru}
\altaffiliation{Facility for Rare Isotope Beams, East Lansing, Michigan 48824, USA}
\affiliation{Department of Physics \& Astronomy,
Ohio University, Athens, Ohio 45701, USA}
\author{A. Parikh}
\affiliation{Department de F\'isica, Universitat Polit\`ecnica de Catalunya, E-08036 Barcelona, Spain}
\author{C. Ruiz}
\affiliation{TRIUMF, 4004 Wesbrook Mall, Vancouver,
British Columbia V6T 2A3, Canada}
\affiliation{Department of Physics \& Astronomy, University of Victoria,
Victoria, BC  V8W 2Y2, Canada}
\author{A.C. Shotter}
\affiliation{School of Physics, University of Edinburgh EH9
3JZ Edinburgh, United Kingdom}
\author{M. Williams}
\affiliation{TRIUMF, 4004 Wesbrook Mall, Vancouver,
British Columbia V6T 2A3, Canada}
\affiliation{Department of Physics, University of York,
Heslington, York YO10 5DD, United Kingdom}

\begin{abstract}
We have performed the first direct measurement of two resonances of
the \beag~reaction with unknown
strengths using an intense radioactive $\isotope[7][]{Be}$ beam and
the DRAGON recoil separator. We report on the first measurement of the
1155 and 1110~keV resonance strengths of $1.73 \pm 0.25(stat.) \pm 0.40(syst.)$~eV and $125 ^{+27}_{-25}(stat.) \pm 15(syst.)$~meV, respectively.
The present results have reduced the uncertainty in the
\beag~reaction rate to $\sim 9.4-10.7$\% over T = 1.5-3~GK, which is relevant for
nucleosynthesis in the neutrino--driven outflows of core--collapse supernovae
($\nu p$--process). We find no effect of the new, constrained reaction rate on $\nu p$--process nucleosynthesis.
\end{abstract}


\maketitle


Nucleosynthesis in the neutrino--driven winds of
core--collapse supernovae (ccSNe) has gained attention
in recent years. The most recent multi--dimensional
hydrodynamic studies of neutrino--driven explosions with
an energy--dependent neutrino transport mechanism suggest
that the early supernova ejecta are proton--rich
(with electron fraction $Y_e \equiv n_p (n_p+n_n)^{-1} >0.5$, where $n_p$ and $n_n$ are the number densities of protons and neutrons, respectively)~\cite{burrows2021core,muller2016status,wanajo2018nucleosynthesis,vartanyan2019successful}. At later times, the wind becomes slightly neutron--rich
($Y_e \sim 0.40-0.49$) and in these conditions the weak
\emph{r}--process produces nuclei up to A$\approx 90-110$,
below the second \emph{r}--process peak~\citep{qian2007oh, arcones2011production, wanajo2013r}.

In the proton--rich environment of the neutrino--driven ejecta,
the $\nu p$--process operates, synthesizing heavy nuclei with
$A > 74$~\cite{frohlich2006neutrino,pruet2006nucleosynthesis,wanajo2006rp}.
At first, the ejected material from the proto--neutron star (PNS)
is very hot and consists mainly of protons and neutrons, with an excess of the
former, since $Y_e > 0.5$.
Expansion causes the ejecta to cool down and $Z=N$ nuclei are assembled
-- mainly $\isotope[56][]{Ni}$ and $\isotope[4][]{He}$ -- via the Nuclear
Statistical Equilibrium (NSE). At T $\sim$ 3~GK, the excess of protons interacts
with the electron antineutrinos that are streaming from the PNS, producing a small amount of
neutrons, which can be immediately captured by $\isotope[56][]{Ni}$. By a series of
$(n,p)$ and ($p,\gamma$) reactions, the reaction flow proceeds to heavier nuclei, until
the ejecta temperature falls to T $\sim$ 1.5~GK, where the ($p,\gamma$) reactions
freeze--out due to the Coulomb barrier.

The aforementioned scenario has been proposed as a possible production mechanism for the
light \emph{p}--nuclei, a subset of the around 35 neutron deficient nuclei with
$A \geq 74$, which cannot be synthesized by either the \emph{s}-- or
the \emph{r}--process~\cite{arnould2003p,rauscher2013constraining}.
In particular, $\isotope[92,94][]{Mo}$ and $\isotope[96,98][]{Ru}$
that are underproduced in the astrophysical
$\gamma$--process~\cite{pignatari2016production},
could be synthesized via the $\nu p$--process. Furthermore, the $\nu p$--process
could also explain the high abundance of Sr, Y and Zr relative to Ba in
metal--poor stars and has been proposed as a candidate of the
light--element primary process (LEPP)
~\cite{montes2007nucleosynthesis, arcones2011production}.

Despite its successes, the $\nu p$--process exhibits many
uncertainties that have already been identified since it
was first proposed. Its efficiency strongly depends on the
characteristics of the neutrino--driven wind (\emph{e.g.} electron
fraction $Y_e$ and entropy $s$) and the underlying nuclear
physics input (\emph{e.g.} reaction rates and $Q$ values)
~\cite{wanajo2011uncertainties, arcones2011production, nishimura2019uncertainties}.

One of the most important reactions affecting
the nucleosynthesis output of the $\nu p$--process
is the triple--$\alpha$ reaction, which controls the relative abundances
of protons, $\alpha$--particles, and $\isotope[56][]{Ni}$ seed nuclei
before the onset and during the $\nu p$--processing~\cite{nishimura2019uncertainties}.
In particular, a high rate of the triple--$\alpha$
reaction decreases the efficiency of the $\nu p$--process
since it creates more seed nuclei, acting as a ``proton poison'' by decreasing
the ratio of neutrons to seed nuclei, $\Delta_n$.
However, \citet{wanajo2011uncertainties} identified a couple
of two--body breakout reaction sequences between $A<12$
(\emph{pp}-chain region) and $A \geq 12$ (CNO region) that can have
a similar effect to the triple--$\alpha$ reaction and compete
with it in the temperature range of the $\nu p$--process,
namely $\isotope[7][]{Be}(\alpha,\gamma)\isotope[11][]{C}(\alpha,p)\isotope[14][]{N}$ and
$\isotope[7][]{Be}(\alpha,p)\isotope[10][]{B}(\alpha,p)\isotope[13][]{C}$.
The most important reaction for each sequence is
\beag~and $\isotope[10][]{B}(\alpha,p)\isotope[13][]{C}$ respectively,
and for this reason they were included in a nucleosynthesis sensitivity study by~\citet{wanajo2011uncertainties}.
Their results suggest that species with $90<A<110$ are
sensitive to variations of the \beag~
reaction rate. Their abundances can vary up to an order of magnitude
when varying the \beag~reaction rate by factors between 0.1 and 10,
and for this reason it needs to be well constrained experimentally.

In the relevant energy region for $\nu p$--process nucleosynthesis
there are three experimental studies of the \beag~
reaction~\cite{hardie1984resonant, wiescher1983c, yamaguchi2013alpha}.
The two low--lying resonances at $E_r=$ 561 and 876~keV
were studied by~\citet{hardie1984resonant}
in forward kinematics using a radioactive $\isotope[7][]{Be}$ target and
their strengths were measured. For the $E_r=$ 1110 and 1155~keV resonances~\citet{wiescher1983c}
used the $\isotope[10][]{B}(p,\gamma)\isotope[11][]{C}$
reaction and calculated their $\Gamma_\gamma/\Gamma$ from the cross section
ratio $\sigma_{(p,\gamma)}/\sigma_{(p,\alpha)}$, but their strengths remain
unknown. The most recent relevant study was performed by~\citet{yamaguchi2013alpha}.
The authors performed a $\isotope[7][]{Be} + \alpha$ resonant scattering
and $\mathrm{\isotope[7][]{Be}(\alpha,p)}$ reaction measurement using the
thick--target method in inverse kinematics and
measured the excitation functions for $E_x$ = 8.7--13.0 MeV on $\isotope[11][]{C}$.
Their \emph{R}--matrix analysis revealed a new state at $E_x$ = 8.9~MeV
($E_r=$ 1356~keV) which could have a 10\% contribution to the total
\beag~reaction rate in the relevant energy region. However, the authors
argue that due to their large uncertainty in the low energy region,
this level might be the $E_x$ = 8.699~MeV ($E_r=$ 1155~keV)
state.

The current rate for the \beag~reaction
is adopted from NACRE (I and II)~\citep{angulo1999compilation, xu2013nacre} and includes
contributions only from the two low--lying (561 and 876~keV) narrow resonances,
for which experimentally measured strengths exist. In the work of~\citet{angulo1999compilation} (NACRE--I), whose rate was used as a baseline in the sensitivity study of~\citet{wanajo2011uncertainties}, Hauser--Feshbach contributions were added for T$>$2~GK. In the most recent evaluation of the rate by~\citet{xu2013nacre} (NACRE--II), the authors included contributions from four broad resonances at higher energies. \citet{descouvemont19957be}
also suggests that the sub--threshold resonance at $E_x$ = 7.4997~MeV ($E_r= -43.9$~keV) can dominate the reaction rate at low temperatures, below $T \approx 0.3$~GK, which
could impact the destruction of the important radionuclide $\isotope[7][]{Be}$
in astrophysical sites such as classical novae and PopIII stars.
The NACRE--II thermonuclear reaction rate is uncertain by factors of 1.76--1.91 for $T=$~1.5--3~GK~\cite{xu2013nacre}. In addition to that, contributions from higher energy resonances with unknown strengths are expected to influence the reaction rate for $T > 1.5$~GK~\cite{wanajo2011uncertainties}.

In this Letter, we present the first experimental study of the
\beag~reaction in inverse kinematics utilizing an intense $\isotope[7][]{Be}$
radioactive ion beam (RIB) to measure two key resonances at
$E_r$= 1110 and 1155~keV, with unknown strengths, and determine
their contribution to the reaction rate at $\nu p$--process
nucleosynthesis energies. In addition, we re--measured the
$E_r$= 876~keV resonance strength.

The measurements were performed using the DRAGON
(Detector of Recoils and Gammas of Nuclear reactions)
recoil separator~\cite{hutcheon2003dragon} at the ISAC--I
(Isotope Separator and Accelerator) experimental hall of TRIUMF,
Canada's particle accelerator centre in Vancouver, BC, Canada.
Intense beams of $\isotope[7][]{Be}$ ($I \sim 1.3-5.8 \times 10^8$~pps) were produced using the ISOL technique,
by bombarding thick ZrC and graphite targets with 55~$\mu A$ 500~MeV protons from the
{TRIUMF} cyclotron. The $\isotope[7][]{Be}$ content of the beam was enhanced compared to
the main $A= 7$ isobar $\isotope[7][]{Li}$ using the
TRIUMF Resonant Ionization Laser Ion Source (TRILIS)~\cite{lassen2005resonant}.
The radioactive beams were then accelerated through the ISAC--I Radio--Frequency Quadrupole
(RFQ) and Drift--Tube Linac (DTL) to energies, so that each resonance was centered in the
gas target. To ensure a pure RIB, an additional 20~$\mu g$/cm$^2$ carbon stripping
foil was placed upstream of the DTL to select a specific charge state (4$^+$)
to completely eliminate the main isobaric contaminant $\isotope[7][]{Li}$.
Finally, $\isotope[7][]{Be^{4+}}$ was delivered to the helium--filled
DRAGON windowless gas target with effective length of 12.3(1)~cm~\cite{hutcheon2003dragon}.
In Table~\ref{tab:1} we present an overview of the beam and gas target parameters for our measurements.

\begin{table}[ht!]
\centering
\caption{Beam and gas target properties for the two independent measurements of
the present study\footnote{The 1110 keV resonance was studied in two independent
measurements, due to a low recoil yield in the first measurement. We quote the weighted average values for the presented quantities.}.}
     \begin{tabular}{cccccc}
     \hline \hline
   $\mathrm{E_{beam}}$   & $\mathrm{E_{lab}}$  & $\mathrm{P_{target}}$  & $\mathrm{E_{c.m.}}$   & $\mathrm{t_{irrad.}}$ & $\mathrm{N_{\isotope[7][]{Be}}}$ \\
    ($A$ keV) & (MeV) & (Torr) & (MeV) & (h) &($\times 10^{13}$) \\ \hline
    464.2(3) & 3.249(2) & 7.9(1)  & $1157 \pm 24$  & 25.4 & 1.07(2)  \\
    442.2(2) & 3.098(1) & 4.92(7) & $1111 \pm 13$  & 34.2 & 3.29(5)  \\
    351.8(3) & 2.463(2) & 5.75(4) & $878 \pm 17$   & 27.8 & 2.12(4)  \\
    \hline \hline
    \end{tabular}
\label{tab:1}
\end{table}

\begin{figure*}[ht!]
    \centering
    \includegraphics[width=\textwidth]{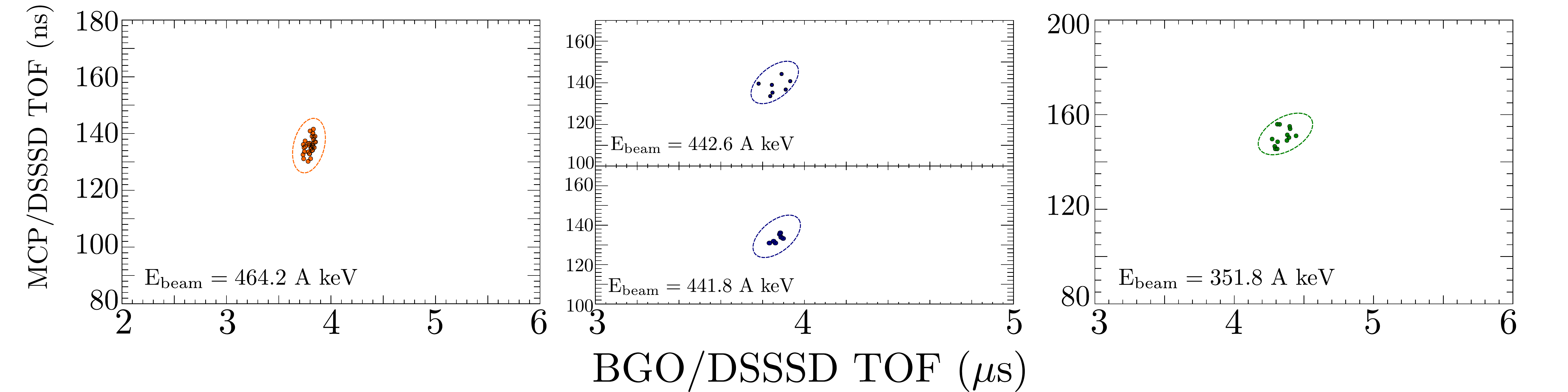}
    \caption{MCP/DSSSD versus BGO/DSSSD (Separator) Time--Of--Flight for the recoil events for
    each of the resonances we studied in the present work. For the 1110~keV resonance, we show the two independent measurements in separate panels. Positively identified $\isotope[11][]{C}$ recoils for each resonance are shown. The ovals are used to help the reader's eye.}
\label{fig:1}
\end{figure*}

An array of 30 highly efficient bismuth germanate (BGO) detectors
surrounding the gas target detected the prompt $\gamma$ rays of the
$\isotope[11][]{C}$
recoil de--excitation and provided $\gamma$ tagging for the coincidence
analysis. The most intense charge state of the recoils ($\isotope[11][]{C}^{2+}$)
was tuned through the separator to a 66~$\mu m$ thick, gridded Double--Sided
Silicon Strip Detector ({DSSSD}) -- Micron W1(G) model -- placed near the focal plane of {DRAGON},
with a typical rate of 5--15~Hz. The $\isotope[11][]{C}$ recoils were detected
both in singles and coincidences modes. In the former, we employed
Time--of--Flight (TOF) measurements between a microchannel plate detector (MCP)
close to the DRAGON focal plane
and the DSSSD, and in the latter we used the detected $\gamma$ rays in the
BGO array and hits on the DSSSD (see the PID plot in Figure~\ref{fig:1}).

\begin{figure}[hbpt!]
    \centering
    \includegraphics[width=.5\textwidth]{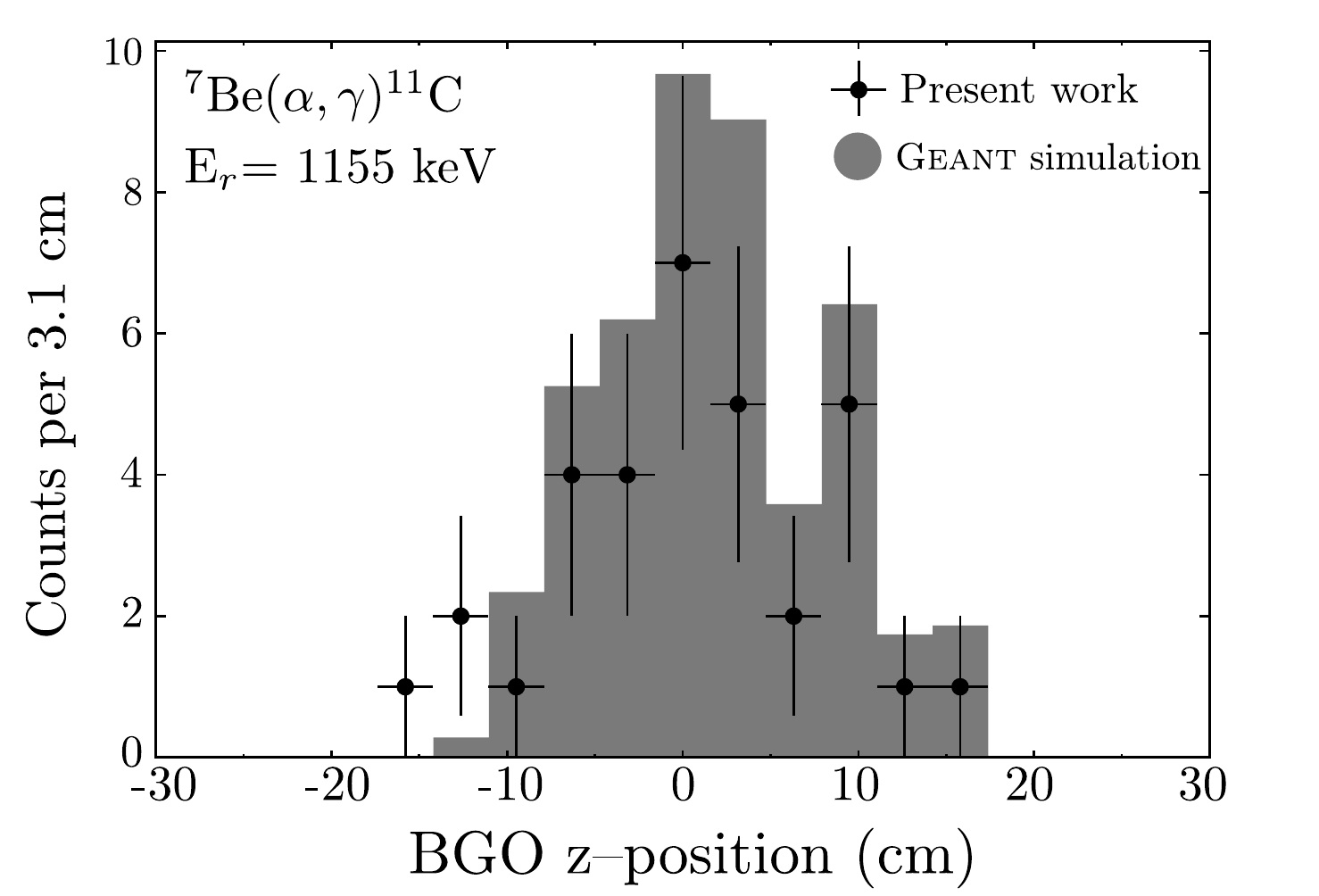}
    \caption{BGO position profile
spectrum for the $E_r$=1155~keV resonance. The black points indicate the experimental data and the gray histogram a \textit{scaled} {\scshape Geant} simulation. The centroid of the experimental peak is at $z_r$= +0.47~cm with respect to the center of the gas target.}
\label{fig:2}
\end{figure}

According to the reaction kinematics for the
\beag~reaction in the energy range of interest, the recoil
angular distribution greatly exceeds the nominal DRAGON acceptance
($\theta_{r,max} \sim 43-47$~mrad compared to $\theta_{DRAGON}$= 21~mrad).
For this reason, we performed detailed simulations using the
standard {DRAGON} {\scshape Geant} package\footnote{The
{\scshape Geant} simulation package of DRAGON can be found at
\href{https://github.com/DRAGON-Collaboration/G3\_DRAGON}{https://github.com/DRAGON-Collaboration/G3\_DRAGON}.}~\cite{gigliotti2003calibration} to
calculate the efficiency of the BGO array ($\eta_{BGO}$) and the transmission of the
recoils through the separator ($\mathrm{\eta_{separator}}$), which are used in the data analysis and the
calculation of the resonance strengths.
This procedure has already been employed successfully in DRAGON experiments
and more recently with a benchmark measurement of a resonance with a known strength of the
$\isotope[6][]{Li}(\alpha, \gamma)\isotope[10][]{B}$ reaction,
whose products also had a maximum angular cone larger than DRAGON nominal
acceptance~\cite{psaltis6li, psaltis2020radiative}. A more detailed discussion
about these simulations can be found in the accompanying publication~\cite{psaltis_prc}.
We observe a very good agreement between the {\scshape Geant} simulations and the
experimental data. In particular, Figure~\ref{fig:2} shows a spectrum of the highest energy
$\gamma$ ray per coincident event versus the position along the beam axis
for the 1155~keV resonance.

The number of the incident beam particles was determined by
using the elastically scattered target particles, using two
silicon surface barrier (SSB) detectors placed at well--defined lab angles
of 30$^\circ$ and 57$^\circ$ with respect to the beam axis. The beam stopping power
through helium gas and the recoil charge state distribution, which are used for
the calculation of the experimental resonance strength, were measured
using $\isotope[7][]{Be}$ and $\isotope[12][]{C}$ beams respectively.

Figure~\ref{fig:1} shows the MCP--DSSSD versus separator TOF for the three resonances
studied in the present work. Clusters of 33, 9/7
and 13 coincidence events were recorded for the 1155, 1110, and 876~keV resonances,
respectively. For all the resonance strength
measurements a very high beam suppression is demonstrated, consistent with the reported
$10^{13}$ in Ref.~\cite{sjue2013beam}.

The main sources of systematic uncertainty in the final result arise from the
$\gamma$ decay branching ratio uncertainties and the $\gamma$ ray angular distributions
which both affect the BGO efficiency, and subsequently the recoil transmission through the
separator~\cite{ruiz2014recoil}. The relative uncertainties of the product of $\eta_{BGO}$ and
$\mathrm{\eta_{separator}}$ uncertainty for the three resonances are the following:
19.9\% ($E_r=$ 1155~keV), 11.0\% ($E_r=$ 1110~keV) and 29.3\% ($E_r$= 876~keV). Smaller
contributions to the systematic uncertainty arise from the MCP detection efficiency
(5.5--10.7\%) and the stopping power measurements (3.7--4.3\%)~\citep[see also the discussion and Table VIII in Ref.][]{psaltis_prc}. The statistical
uncertainties in turn are due to the low detection yield, caused by
the very low transmission of the recoils through the separator. However, even though the transmission is small, it is a parameter that is well understood and quantified~\cite{psaltis6li}. The detected recoil uncertainties for the 1110 and 876~keV resonances were determined
using the prescription of~\citet{feldman1998unified} for a poissonian signal with zero
background, as it is evident in Figure~\ref{fig:1}.

We determined the resonance strengths of the 1155, 1110 and 876~keV resonances to
be $1.73 \pm 0.25(stat.) \pm 0.40(syst.)$~eV, $125 ^{+27}_{-25}(stat.) \pm 15(syst.)$~meV and $3.00^{+0.81}_{-0.72} (stat.) \pm 0.61(syst.)$~eV,  respectively. For the 1110~keV
resonance strength, since we performed two independent measurements, we created a combined statistical uncertainty distribution, accounting for the
asymmetric statistical uncertainties from the prescription of~\citet{feldman1998unified}. We provide a detailed discussion of this procedure in ~\citet{psaltis_prc}.

\begin{figure}
\centering
        \centering
    \includegraphics[width=.5\textwidth]{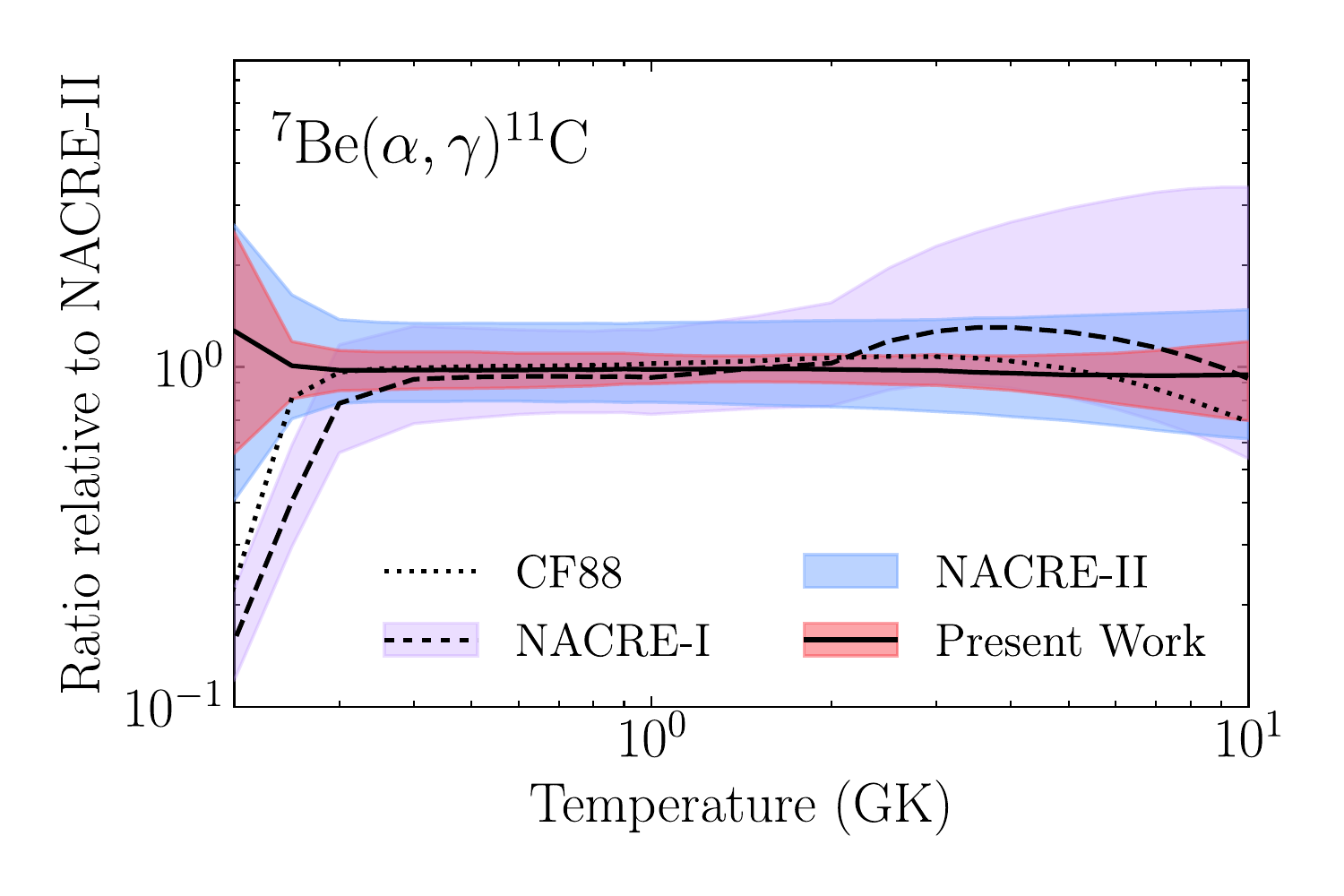}
    \caption{Comparison between the NACRE-I~\citep{angulo1999compilation}, NACRE--II~\citep{xu2013nacre} $\isotope[7][]{Be}(\alpha,\gamma)\isotope[11][]{C}$ thermonuclear reaction rate and that of~\citet{caughlan1988thermonuclear} (CF88) and the present work for the same temperature region.}
        \label{fig:rate}
\end{figure}

Using the results from the present experiment, we evaluated the \beag~reaction rate using the RatesMC code\footnote{The RatesMC code to calculate thermonuclear reaction rates can be found at \href{https://github.com/rlongland/RatesMC}{https://github.com/rlongland/RatesMC}.}~\cite{longland2010charged}.
Figure~\ref{fig:rate} shows the new reaction rate in comparison with both the NACRE rates~\cite{angulo1999compilation,xu2013nacre} and the compilation of~\citet{caughlan1988thermonuclear} (CF88).
The new \beag~reaction rate differs less than $\sim 2\%$ at temperatures between T= 1.5-3~GK with the NACRE-II rate, but it is now constrained to $\sim 9.4-10.7$\%, which is sufficient for astrophysical applications. It is worth noting that the decrease in the rate uncertainty mainly originates from properly propagating the relevant errors within the RatesMC framework, and the newly measured resonance strengths contribute $\lessapprox 10$\% to the total rate in $\nu p$--process temperatures~\citep[see the discussion in Ref.][Sec. E]{psaltis_prc}.

Furthermore, we performed nucleosynthesis calculations using the new \beag~reaction rate and parametric neutrino--driven wind trajectories from Ref.~\cite{jacobimsc} to study the impact on the production of heavy elements.
Despite the fact that the new \beag~thermonuclear reaction rate is more constrained compared to NACRE--II, we did not observe any differences in the production of heavy elements via the $\nu p$--process. For completeness, and to note the sensitivity of $\nu p$--process nucleosynthesis to this rate, we did find that a \beag~reaction rate $\sim 2$ times lower than the NACRE--II increased the production of A = $55 - 130$ nuclei by as much as a factor of 100 in specific astrophysical conditions of the neutrino--driven wind~\cite{jacobimsc}, which is significantly larger than the typical abundance changes observed by~\citet{wanajo2011uncertainties}. Such a rate reduction is beyond our determined rate uncertainty, however. That said, a future detailed study of this nucleosynthesis scenario will examine if such discrepancies also exist for other important nup-process rates and include the results from recent measurements of such reactions~\cite{liu2020low, kibedi2020, randhawa2021}.

In addition to the nuclear physics uncertainties, the $\nu p$-process is strongly dependent on the local astrophysical conditions of the neutrino--driven wind and more specifically on the combination of $Y_e$, $s$ and expansion timescale $\tau$. Given that the state--of--the--art multi--dimensional simulations of ccSNe support proton--rich outflows~\cite{burrows2021core,muller2016status,wanajo2018nucleosynthesis,vartanyan2019successful}, the $\nu p$--process should be a very common nucleosynthesis scenario and its yields need to be included in Galactical Chemical Evolution (GCE) models. Nevertheless, as~\citet{kobayashi2020} have argued, the inclusion of such yields leads to an overproduction for elements between strontium (Sr) and tin (Sn), compared to observations. For this reason, we argue for a coordinated effort between experimental nuclear physicists, stellar modellers and observational astronomers to constrain the most common conditions for the $\nu p$--process and its role in the origin of the heavy elements in the universe.

To recapitulate, in this Letter we presented the first inverse
kinematics study of the \beag~reaction using the DRAGON recoil separator
and an intense $\isotope[7][]Be$ beam from ISAC. We successfully measured for the
first time the strength of two resonances
at 1155 and 1110~keV ($\omega \gamma= 1.73 \pm 0.25(stat.) \pm 0.40(syst.)$~eV and $125 ^{+27}_{-25}(stat.) \pm 15(syst.)$~meV), and remeasured one at 876~keV ($\omega \gamma=$ $3.00^{+0.81}_{-0.72} (stat.) \pm 0.61(syst.)$~eV), which agrees within uncertainty with the measurement of~\citet{hardie1984resonant} ($\omega \gamma= 3.80(57)$~eV).
The uncertainty of the \beag~reaction rate in now reduced to $\sim 9.4-10.7$\% at
the temperature region relevant to $\nu p$--process nucleosynthesis, T= 1.5-3~GK. According to
our results, the new reaction rate is well constrained for astrophysical calculations,
and our initial nucleosynthesis calculations suggest that it does not affect the production of
neutron--deficient heavy elements (\textit{p}--nuclei).
This experiment is a major technical achievement, being the first radiative capture reaction measurement using a RIB and a recoil separator, in which the angular distribution of the reaction products exceeds the nominal acceptance of the separator by more than a factor of two.
In addition, the intense $\isotope[7][]{Be}$ radioactive beams
produced with the use of graphite targets,
can be employed for other challenging measurements, such as the
$\isotope[7][]{Be}(p,\gamma)\isotope[8][]{B}$ and
$\isotope[7][]{Be}(\alpha,\alpha)\isotope[7][]{Be}$ reactions.

The authors gratefully acknowledge the beam delivery and
ISAC operations groups at TRIUMF. In particular, we thank
F. Ames, T. Angus, A. Gottberg, S. Kiy, J. Lassen
and O. Shelbaya for all their help during the experiment. The core operations of TRIUMF
are supported via a contribution from the federal government through the
National Research Council of Canada, and the Government of British
Columbia provides building capital funds. Authors from
McMaster University are supported by the National Sciences
and Engineering Research Council of Canada (NSERC). DRAGON is funded through
NSERC grant SAPPJ-2019-00039. AP also acknowledges support from the Deutsche Forschungsgemeinschaft (DFG, German Research Foundation)-Project No. 279384907-SFB 1245, and the State of Hesse within the Research Cluster ELEMENTS (Project ID 500/10.006).
RL and CM acknowledge support from the
U.S. Department of Energy, Office of Science, Office of Nuclear Physics under
Grant Nos. DE-SC0017799 and DE-FG02-97ER41042.
CRB, RG and SP acknowledge support from the U.S. Department of Energy, under grants number DE-FG02-88ER40387 and DE-NA0003883.
Authors from the Colorado School of Mines acknowledge support from U.S. Department of
Energy Office of Science DE-FG02-93ER40789 grant.
This work benefited from discussions at the ``Nuclear Astrophysics at Rings and Recoil Separators" Workshop
supported by the National Science Foundation under Grant No.
PHY-1430152 (JINA Center for the Evolution of the Elements).

\bibliographystyle{apsrev4-2}
\bibliography{apssamp}

\end{document}